\documentclass[a4paper]{article}
\usepackage{epsfig}
\usepackage{amssymb}
\usepackage{color}
\normalsize
\newcommand{\de}{\delta}
\newcommand{\De}{\Delta}

\newcommand{\ga}{\gamma}

\newcommand{\La}{\Lambda}

\newcommand{\Om}{\Omega}

\newcommand{\si}{\sigma}

\newcommand{\ra}{\rightarrow}

\newcommand{\be}{\begin{equation}}
\newcommand{\ee}{\end{equation}}
\newcommand{\gsim}{\stackrel{>}{\sim}}
\newcommand{\lsim}{\stackrel{<}{\sim}}
\newcommand{\bea}{\begin{eqnarray}}
\newcommand{\eea}{\end{eqnarray}}
\newcommand{\bean}{\begin{eqnarray*}}
\newcommand{\eean}{\end{eqnarray*}}
\newcommand{\dd}{\partial}

\newcommand{\LCDM}{$\Lambda$\rm{CDM}}

\definecolor{darkred}{rgb}{1,0,0.2}
\definecolor{back}{rgb}{0.8,0.95,0.8}

\begin{document}
\title{What do we really know about Dark Energy?}

\author{Ruth Durrer\footnote{ruth.durrer@physics.unige.ch}\\
CEA, Institut de Physique Th\'eorique, 91191 Gif-sur-Yvette, France\\ 
CNRS, URA-2306, 91191 Gif-sur-Yvette, France\\ and \\
Universit\'e de Gen\`eve, DPT and CAP, 1211 Gen\`eve, Suisse }

\maketitle


\begin{abstract}
In this paper we discuss what we truly know about dark energy. I shall argue that up to date
our single indication for the existence of dark energy comes from distance measurements
and their relation to redshift. Supernovae, cosmic microwave background anisotropies and observations
of baryon acoustic oscillations, they all simply tell us that the observed distance to a 
given redshift $z$ is larger than the one expected from a Friedmann Lem\^\i tre 
universe with matter only and the locally measured Hubble parameter. 
\end{abstract}

\section{Introduction}
\label{sec:intro}
Nearly thirteen years ago, measurements of the luminosity of type Ia supernovae
(SN1a) as function of their redshift~\cite{SNorig} 
have led to the interpretation that the expansion our Universe is presently accelerated and
therefore the energy density of the Universe is presently dominated by a component with 
strongly negative pressure, $P<-\rho/3$, like during inflation. This was an entirely unexpected result
but it has been confirmed with many more observations from SN1a data~\cite{SNnewest}, from
observations of cosmic microwave background (CMB) anisotropies and 
polarization~\cite{wmap-etal}, from weak 
lensing~\cite{weakLensing}, from baryon acoustic oscillations (BAO)~\cite{BAO}, from 
galaxy surveys~\cite{SLOAN} and from cluster data~\cite{cluster-obs}. All this data is
consistent with the so called concordance model, a Friedmann Lema\^\i tre (FL) universe 
with a nearly scale invariant spectrum of Gaussian  initial fluctuations as
predicted by inflation. 

In the concordance model, the energy content of the Universe is dominated
by a cosmological constant $\La \simeq 1.7\times 10^{-66}($eV$)^2$ such that 
$\Om_\La = \La/(3H_0^2) \simeq 0.73$. Here $H_0$ denotes the Hubble constant that we
parameterize as $H_0 = 100 h$km~s$^{-1}$Mpc$^{-1} = 2.1332 h \times 10^{-33}$eV. The second
component of the concordance model is pressure-less matter with $\Om_m = \rho_m/\rho_c =
\rho_m/(3H_0^2/8\pi G) \simeq 0.13/h^2$.  Here $G$ is Newton's constant. About 83\% of 
this matter is 'dark matter', i.e. an unknown non-baryonic component (termed CDM for 'cold 
dark matter') and only about 17\% 
is in the form of baryons (mainly hydrogen and helium), $\Om_bh^2 \simeq 0.022$.
The energy densities of photons and neutrinos are subdominant, 
$\Om_\ga h^2 =2.48\times 10^{-5}$, $0.002<\Om_\nu h^2<0.01$, and 
curvature is compatible with zero.

This situation is disturbing for two main reasons:
\begin{enumerate} 
\item The two most abundant components of the Universe have only been inferred by their
gravitational interaction on cosmological scales.\\
{\bf Dark matter:} on the scale of galaxies, clusters and the Hubble scale. \\
{\bf Dark energy:} only on the Hubble scale. 
\item Including particle physics into the picture, we realize that the cosmological 
constant is in no way distinguishable from vacuum energy. The latter has not
only also the form $T_{\mu\nu}^{\rm vac} = \rho^{\rm vac}g_{\mu\nu}$, but it also 
couples only to gravity.
Hence there is no experiment that can ever distinguish between a cosmological constant $\La$
and a vacuum energy density\footnote{Differences of vacuum 
energies are of course very well measurable e.g. via the Casimir force or the Lamb shift 
in atomic  spectra.}    $\rho_{\rm vac}= \La /(8\pi G)$. My conclusion is, that we therefore 
should not distinguish between the two. We then find that cosmology determines the present 
vacuum energy density to be $\rho_{\rm vac}\simeq (2.7\times 10^{-3}$eV$)^4h^2$.
On the other hand, 'natural values' for the vacuum energy are, e.g. the supersymmetry breaking scale that must be larger than about 1TeV or, if there is no supersymmetry at this scale, the string scale or the Planck scale. The resulting estimates for $\rho_{\rm vac}$ are by 60 respectively 120 orders of magnitude too large. Probably the worst estimate ever in physics! 
Of course, we can introduce a counter term to compensate the 'bare vacuum energy' 
in order to obtain the true, observed value. But, unlike, e.g. the electric charge, vacuum 
energy density is not protected in quantum theory. Corrections to it run like $E_{\max}^4$ 
where  $E_{\max}$ is the cutoff of the theory. Hence the tiny value of $\rho_{\rm vac}$ 
has to be readjusted at each order in perturbation theory by a corresponding, much 
larger counter term. A truly unsatisfactory situation.
Even if we are ready to accept this and say that this is an UV problem with quantum 
field theory that should not be mixed up with the IR problem of the cosmological constant, 
in principle, we also have to introduce a time dependent  IR cutoff to the vacuum energy 
we expect to run like $H(t)^4$, where $H(t)$ is the Hubble parameter at time $t$. Such 
a contribution that at present has to be of  the order of $3H_0^2/(8\pi G)$ was much
larger in the past and is clearly in
 contradiction with cosmological observations. \\ See~\cite{carlo} for the opposite point of view
 on the dark energy problem.
\end{enumerate}
 Dark matter cannot be any particle of
the standard model since all stable standard model particles except the 
neutrino and the graviton, either 
emit photons or would have left their imprint on nucleosynthesis (baryons). Neutrinos, on the 
other hand, have too small masses, too large free-streaming scales, to  
account for the dark matter seen e.g. in dwarf galaxies~\cite{nuDwarfs}
and for other aspects of clustering on small scales (e.g Ly-$\alpha$~\cite{nuLya}). 
This is even more true for the graviton which is massless. But we know  from particle physics
that there have to be modifications to the standard model at energies not much 
larger than 1TeV. Most of the popular proposals of such modifications, like e.g. supersymmetry,
do predict massive stable particles with weak interaction cross sections and the correct
abundance so that they  could play
the role of dark matter. Hence there is no shortage of very reasonable candidates 
which we have not been able to detect so far. Furthermore, if e.g. the simplest
supersymmetric models are realized, and the dark matter particle is the neutralino, 
there is justified hope to detect it soon, either at LHC (Large Hadron Collider at 
CERN)~\cite{LHCdm} or via direct dark matter 
detection experiments~\cite{dmdirect}.

Dark energy, however, is very disturbing. On the one hand, the fact that such an
unexpected result has been found by observations shows that present cosmology
is truly data driven and not dominated by ideas which can be made to fit sparse observations.
Present cosmological data are too good to be brought into agreement with vague
ideas. On the other hand, a small cosmological constant is so unexpected and so difficult
to bring into agreement with our ideas about fundamental physics, that people have
started to look into other possibilities. 

One idea is that the cosmological constant 
should be replaced by some other form of 'dark energy', maybe an ordinary or
a tachyonic scalar field~\cite{quintessence}.
Another possibility is to modify the left hand side of Einstein's equation, i.e.
to modify  gravity. For a review see~\cite{RR12}. Models where the Einstein-Hilbert action
is modified by $R \ra f(R)$ have been investigated~\cite{Sergio}. Another direction
are theories with extra
dimensions (for reviews/introductions see e.g.~\cite{branes}) which, when reduced to 
four dimensions contain terms which deviate from Einstein
gravity, the simplest and best studied example is the DGP model~\cite{DGP}.

All these models are quite speculative and one has to test in each case that they do not contain dangerous 'ghosts'  or other instabilities, that are expected from generic higher derivative 
terms~\cite{ostro,woodard}. Furthermore, even if one of these models is realized in nature, 
the question, why we do {\em not} measure a cosmological constant gravitationally 
remains. However, there may by
more satisfactory ways to address this question, like de-gravitation~\cite{degrav} or emerging
gravity~\cite{Jacobson,emergePaddy}.

Another, more conservative possibility is to take into account the fact that the true Universe is 
inhomogeneous at least on small scales. The question then is wether the clumpiness 
could mimic the presence of a cosmological constant~\cite{bucher,syksy,chrisroy}. 
Another, more extreme attempt is to assume that the background universe 
is not homogeneous but only isotropic, a Lema\^\i tre-Tolman-Bondi model~\cite{Enqvist}. 
Interestingly, these questions are still open. I shall come back to this point later.

In the present paper I do not want to discuss or judge these possibilities, but I want to 
investigate what present data really has measured. As always when our interpretation 
of the data leads us to a very unexpected, unnatural 'corner' in the space of physical 
theories, it may be useful to take a step back and reflect on what the measurements 
really tell us and how much of what we conclude is actually an {\em interpretation} of the data 
that might be doubted.

In the next section I shall go over the main physical observations one by one and 
address this question. In section~\ref{s:whatDE} we discuss what this means for dark energy 
and in section~\ref{s:conc} I conclude.

\noindent
{\bf Notation}: We use $t$ as conformal time such that $ds^2 =a^2(t)(-dt^2 +\ga_{ij}dx^idx^j)$.
The scale factor is normalized to one today, $a_0 =a(t_0)=1$ but spatial curvature $K$ is 
arbitrary. $\Om_X = \rho_X(t_0)/\rho_c(t_0) =  8\pi G\rho_X(t_0)/(3H_0)^2$ is the (present)
density parameter of the component $X$.

\section{What do we really measure?}
\subsection{Supernovae Ia}
Let us start with the first data that gave strong indication of an accelerating universe, the 
supernovae type Ia observations. SN1a observations measure the light curve
and the spectrum of supernovae. The latter is not only used to determine the redshift,
but also indicative for the type of the supernovae while 
the light curve can be translated into a luminosity distance, $D_L(z)$ to the supernova.
For this, correlations between the light curve maximum and its width are used to reduce 
the scatter and derive the intrinsic luminosity. SN1a are so called modified standard 
candles~\cite{SNnewest}. By this correction, the intrinsic scatter of SNIa luminosities of about
$1.5$mag can be reduced to 0.2mag~\cite{kim}. It is very likely that in the near future this
error can be reduced by at least a factor of two~\cite{bob}.  Astronomical magnitudes are 
related to the luminosity distance by
\be
m(z_1)-m(z_2) = 5\log_{10}\left(D_L(z_1)/D_L(z_2)\right) \,.
\ee
Hence an error in the magnitude translates to an error in the luminosity distance via
\be
\frac{\de D_L(z)}{D_L(z)} = \frac{\log(10)}{5}\de m(z)  = 0.46\de m(z).
\ee
Or, an error of 0.2 in the magnitude corresponds to an error of nearly 10\% in the 
luminosity distance. This is the optical precision we can reach at this time, not including
systematic errors like e.g. evolution. 

If we now {\em assume} that the geometry of the Universe is Friedmann Lema\^\i tre (FL), 
we can relate the luminosity to the energy content of the universe via the standard formula
\bea 
D_L(z) &=& (1+z)\chi_K\left(\int_0^z \frac{dz'}{H(z')}\right) \quad \mbox{ where } ~
\chi_K(r) =  \frac{1}{\sqrt{K}}\sin(r\sqrt{K}) \,, ~  \mbox{ and} \nonumber\\   \label{e:DL}
H(z) &=& H_0\sqrt{\Om_m(1+z)^3 +\Om_K(1+z)^2 +\Om_r(1+z)^4 
+\Om_{DE}(z) } \,.
\eea
Here $K$ is spatial curvature and $\chi_K(r) \ra r$ for $K\ra 0$. For negative values of $K$ the
square roots become imaginary and $\sin(r\sqrt{K})/\sqrt{K} = \sinh(r\sqrt{|K|})/\sqrt{|K|}$. 
$\Om_K =-K/H_0^2$ and $\Om_{m,r}$ are the matter respectively radiation density parameter.
$\Om_{DE}(z) = \rho_{DE}(z)/\rho_c(z=0)$ is the contribution from dark energy.
For a cosmological constant $\Om_{DE}(z) =\La/(3H_0^2)$ is constant.

\begin{figure}
\centering  
\includegraphics[width=8cm]{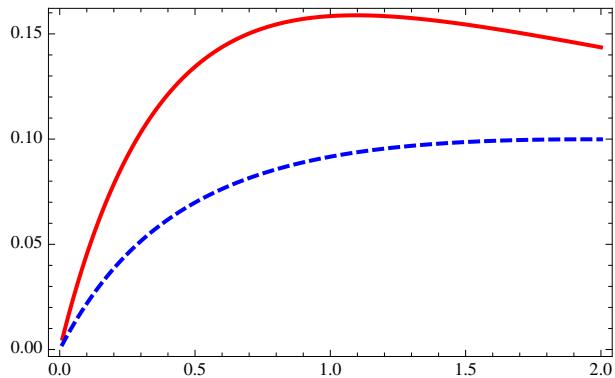}
\caption{ The relative differences $(D_L - D_L(0.7,0.3,0))/D_L) $ (blue, solid)  and $(D_L - D_L(0,0,1))/D_L) $ (red, dashed)are shown as function of the redshift $z$, where $D_L$ is the luminosity distance
 for a pure CDM model $(\Om_\La, \Om_m,\Om_K) =(0,1,0)$. \label{f:DL}}
\end{figure}

In Fig.~\ref{f:DL} we show the relative difference in the luminosity distance of a Universe 
with the density parameters between a pure CDM model with $(\Om_ \La ,\Om_m,\Om_K) =(0,1,0)$
and a concordance model$(\Om_ \La ,\Om_m,\Om_K) =(0.7,0.3,0)$ as well as between   
pure CDM and an open model, $(\Om_ \La ,\Om_m,\Om_K) =(0,0,1)$ .
The first difference is larger than 10\% already for redshifts $z>0.2$ and should therefore 
easily be visible in
present supernova data. The second never gets larger than 0.1, but observations of many
supernovae should still easily distinguish a $\La$-dominated universe from a negative curvature
dominated one. This is what SN1a observers claim they can do. Most of the data comes from redshifts below and up to $z\simeq 1$. In this regime, observers therefore detect a luminosity distance which
is significantly larger than the one of a flat matter dominated or a curvature dominated Universe with the same Hubble constant.

Hence, if the error estimates of SN1a observers  can be trusted, these data indicate 
either that the geometry of the Universe be not Friedmann, or that the
luminosity distance is dominated at low redshift by an accelerating component which behaves
similar to a cosmological constant.

\subsection{Baryon acoustic oscillations}
Another way to measure distances is to compare angles subtended by objects of a given size 
when placing them at different redshifts. For any metric theory, this angular diameter distance 
is simply related to the luminosity distance by
$$ D_A =D_L/(1+z)^2 \, .  $$
Baryon acoustic oscillations are the relics in the matter power spectrum of the 
oscillations in the baryon-photon plasma prior to decoupling. Once hydrogen recombines 
and the photons decouple from the electrons, the baryon
perturbations evolve like the pressure-less dark matter. Matching this evolution to the oscillations
prior to decoupling, one obtains for the positions of the peaks and troughs in the baryon spectrum
\be
k_{n {\rm through}} = \left(n+ 1/2\right)\pi/s   \quad \mbox{ and }~
k_{n {\rm peak}} = \left(n+ 3/2\right)\pi/s \,,
 \ee
 where $s$ is the comoving sound horizon at decoupling,
 \be
  s = \int_0^{t_{\rm dec}} \! c_s dt  \, . 
 \ee
More precise values are obtained by numerical codes like CAMBcode~\cite{CAMBcode} and by
analytical fits~\cite{Eisen,FM}. The angular diameter distance measures a scale
subtended at a right angle to the line of sight. If we can measure the difference in redshift
$\De z$ between the 'point' and the 'tail' of an object aligned with the line of sight, the corresponding
comoving distance  is given by
\be
\De t(z) = \frac{\De z}{H} = \frac{\De z}{1+z}D_{H} (z)~, \quad  D_H = \frac{z+1}{H(z)} \, .
\ee
With present data on large scale structure we have just measured the 3-dimensional 
power spectrum in different redshift bins. We cannot yet distinguish between transverse 
and longitudinal directions. This measures a (comoving) geometrical mean 
$D_{\rm V}(z) = (D_{H} (z)D_{A} (z)^2)^{1/3}$. Results for this scale at redshifts 
z=0.275 and the ratio $D_{\rm V}(z_2)/D_{\rm V}(z_1)$ for $z_2=0.3$ and 
$z_1=0.2$ have been published~\cite{BAO}.

The observational results from the luminous red galaxy sample of the Sloan Digital Sky Survey (SDSS) 
catalog~\cite{BAO} are in good agreement with the cosmological concordance model
$\La$CDM cosmology with $\Om_\La \simeq 0.7$, $\Om_m \simeq 0.3$ and no appreciable 
curvature. Of course, this experiment measures in principle the same quantity as SN1a
observations since for all metric theories of gravity $D_A =D_L/(1+z)^2$, but BAO
observations have very different systematics and the fact that they agree well with SN1a
is highly non-trivial. There are however objections to the significance of the BAO measurements,
see e.g. Refs~ \cite{Sylos,Gasta}.

Again, these data support a measurement of distance which is significantly larger than the
distance to the same redshift with the same Hubble parameter in a $\Om_m =1$ Universe.

\subsection{CMB}
\begin{figure}
\centering  
\includegraphics[width=8cm]{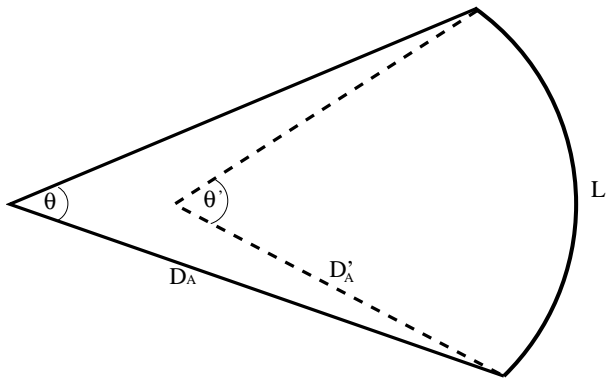}
\caption{ The change of the angle subtended by the CMB acoustic peaks when changing 
only the distance to the last scattering surface. \label{f:angle}}
\end{figure}

Our most precise cosmological measurements are the CMB observations that have 
determined  the CMB anisotropies and polarization to high precision~\cite{wmap-etal}.
These measurements will even be improved substantially by the Planck satellite 
presently taking data~\cite{Planck}. The CMB data is doubly precious since they are not 
only very accurate but also relatively simple to calculate in a perturbed FL universe 
which allows for very precise parameter estimation. For an overview of the physics 
of the CMB, see~\cite{mybook}. The positions of the acoustic peaks, the relics
of the baryon-photon oscillations in the CMB power spectrum, allow
for a very precise determination of the angular diameter distance to the last scattering surface.
If this distance is changed, keeping all other cosmological parameters fixed, the CMB power spectrum,
changes in a very simple way, as shown in Fig.~\ref{f:angle}.

The angle $\theta$ subtended by a given scale $L$ simply changes to  
$\theta' = \theta\cdot (D_A/D_A')$.
Assuming that the signal comes entirely from the last scattering surface and is not 
influenced otherwise by the change of its distance from us (this neglects the 
integrated Sachs Wolfe effect relevant for low harmonics $\ell$),
we can assume that the correlation function of the CMB sky at distance $D_A'$ at angle 
$\theta'$ is equal to that of the CMB sky at distance $D_A$ at angle $\theta$, 
$C'(\theta') = C(\theta)$. Translating this to the power spectrum one obtains for $\ell\gsim 20$,
see Ref.~\cite{MSR},
\be\label{Clrel}
  C_\ell =  \left(\frac{D_A'}{D_A}\right)^2 C'_{\frac{ D_A'}{D_A}\ell}  \, .
\ee
 In addition to this distance which is very well measured 
by  CMB experiments, also the matter and baryon density at the last scattering surface 
as well as the spectral index $n$ and the fluctuation amplitude $A$ are well determined by the CMB. Assuming that dark matter 
and baryons are neither destroyed nor generated between the time of last scattering and today, this leads to the well known value of their present density parameters.

Present CMB data can be fit equally well by the concordance \LCDM model as by a flat matter
dominated model with nearly the same values for $\Om_mh^2$,  $\Om_bh^2$, and $n$
where the angular diameter distance is scaled to a value which is in good agreement with
$D_A$ from the concordance model, $D_A(z_*) \simeq 12.9$Mpc. (Note that this is the 
distance as measured at decoupling, today its value is $(z_*+1) D_A(z_*) $.) More details on 
these results can be found in Ref.~\cite{MSR}.

We have discussed this here not because with think that the true model is actually the 
CDM model with $\Om_m=1$, but  to make clear what aspect of dark energy the CMB 
data really measure:
it is again a distance, the distance to the last scattering surface, i.e. to $z_*\simeq 1090$.

\subsection{Weak lensing}
Weak lensing measurements determine the weak distortion of galaxy shapes by 
gravitational lensing from the matter distribution in the foreground  of the imaged galaxies.
The advantage is that this signal is sensitive only to the total clustered mass in front of
the galaxy. The disadvantage is that the signal is small, the ellipticity due to lensing is 
only about 1\% of  the typical intrinsic ellipticity of galaxies, and only statistical results 
can be obtained. For a review see~\cite{weakLensing}. So far, because of
limitations on the knowledge of the redshift distribution of foreground galaxies and
other statistical problems, weak lensing has mainly
been used to determine the combination $\si_8\sqrt{\Om_m} \simeq 0.6$, which leads 
to 'bananas' in the $\si_8$ -- $\Om_m$ plane. But future surveys like DES (Dark Energy Survey)
or Euclid are 
expected to lead to significant improvements, see Ref.~\cite{wl.fore} for forecasts.
Here $\si_8$ is the amplitude of matter fluctuations in spheres of radius $8h^{-1}$Mpc.
It is determined by the amplitude $A$ of CMB anisotropies and the spectral index $n$.

This measurement by itself may not be so interesting for dark energy, but in combination
with CMB anisotropies it is consistent with the same amplitude, spectral index and, especially matter density as the CMB and therefore can be regarded as independent support of the CMB
result. It is also interesting that this provides a measurement of $\Om_mh^2$ at low 
redshift, $z<1$ which is consistent with the CMB result at $z=z_* \simeq 1090$.

\subsection{Large scale structure}
One of the oldest cosmological measurements are determinations of the correlation 
function and the power spectrum of the galaxy distribution. At present, the biggest 
galaxy survey is the Sloan Digital Sky Survey, SDSS~\cite{SLOAN} which has 
mapped the galaxy distribution on the northern hemisphere out to redshift 
$z\simeq 0.2$ and the luminous red galaxies to $z\simeq 0.5$. This led to a
determination of the galaxy power spectrum down to 
$k\simeq 0.02h/$Mpc~\cite{reid}.

The main problem here is that we compare this measured galaxy power spectrum 
with the calculated matter power spectrum. The latter can be calculated very 
accurately on large scales by relativistic cosmological perturbation theory and quite 
accurately on small scales by Newtonian $N$-body simulation. However, the relation 
between this matter distribution and the distribution of galaxies is still to some
extend an unsolved debate which goes under the name of 'biasing'. On small scales, 
it is clear that the galaxy formation process is highly non-linear and may depend on 
other parameters than the matter density alone (e.g. the metallicity which would favor 
the formation of galaxies in the  vicinity of already existing galaxies). 

On large scales, 
most workers in the field assume that bias is linear and close to one, however, simple
investigations of a toy model biasing scheme show that contrary to the
matter distribution, the galaxy distribution may very well acquire a white noise 
component which would dominate on very large scales~\cite{rms,jr}.

If we disregard these problems and assume that in the measured range, or at least where
linear perturbation theory applies, bias is linear, we can also use the galaxy power 
spectrum to get a handle on $\si_8$, $n$ and $\Om_mh^2$, with different systematics 
than from other probes. Interestingly, the power spectrum bends from $\propto k$
behavior to $\propto k^{-3}\ln^2(kt_{\rm eq}$) behavior at the equality scale,
$k_{\rm eq}\simeq \pi/t_{\rm eq} \propto \Om_mh$ in units of $h/$Mpc, 
since $t_{\rm eq} \propto 1/(\Om_mh^2)$. The position of this
turnover (which is very badly constrained with present data), together with the amplitude 
which is proportional to  $\Om_mh^2$ would allow us to infer both, the Hubble parameter 
and the matter density parameter, from the matter power spectrum.

Features in the galaxy power spectrum, like the BAO's or redshift space distortions
might actually be less affected by biasing and therefore provide more promising 
cosmological probes. However, since they contain less information than the full power
spectrum, measuring the latter will always have an advantage.

Finally, in future surveys which go out to very large scales, $z\simeq 2$,
it will be very important to clearly relate the observed galaxy distribution to
 relativistic linear perturbation variables, i.e. to take into account relativistic
effects in the matter power spectrum~\cite{yoo1,yoo2,cr}. This actually does not only 
represent an additional difficulty but even more a new opportunity.

\subsection{Cluster abundance and evolution}
The earliest data favoring a low density Universe probably comes from the observation of cluster abundance and evolution~\cite{BahCen}. Clusters are the largest bound structures in the 
Universe and as such very sensitive the amplitude of density fluctuations on large scales
$\propto \si_8\Om_m$. Actually, clusters usually form at fixed velocity dispersion. 
Therefore, the cluster density strongly constrains
the velocity power spectrum, $P_V \propto \Om_m^{1.2}\si_8^2$  (see, e.g.~\cite{mybook}). 
 Comparing observations with numerical simulations gives~\cite{Pp} 
 $$ \Om_m^{0.6}\si_8 = 0.495\pm\begin{array}{c} 0.034 \\  0.037\end{array} \, .$$
If we insert $\si_8 \simeq 0.8$, this is in rough agreement with $ \Om_m \simeq 0.3$, but certainly requires $ \Om_m < 1$.


\section{What do we know about 'dark energy'?}\label{s:whatDE}
What do these observations really tell us about dark energy? I think it is clear, even though I did not
enter into any details about observational problems, that each observation taken by itself is not conclusive. There are always many things that can go wrong for any one cosmological probe.
We have assumed that systematics are reasonably well under control and we can trust our results.
This is supported by the fact that many different probes with independent systematics give the
same result:
A value of $\Om_mh^2 \sim 0.13$ and a distance to redshift relation at $z\lsim 1$ that is not in agreement
with  flat matter dominated universe but with a \LCDM~universe. 

However, we do not measure 
$\La$ with any cosmological probe. We only infer it from distance measurements by assuming that
the formula~(\ref{e:DL}) can be applied which only holds for homogeneous and isotropic FL models.
On the other hand, we know that the true Universe is at least perturbed. Naively, one may argue
that the gravitational potential is small $\Psi \sim 10^{-5}$ and therefore corrections coming 
from clustering will be small.  But even if $\Psi$ is small, we know that curvature perturbations
which are second derivatives, $\dd_i\dd_j\Psi \sim 4\pi G\de\rho \gsim  4\pi G\rho$ are not small.
On galactic scales they are many orders of magnitude larger than the background term,
$|\dd_i\dd_j\Psi| \gg H^2$. Since
such terms may well enter into the perturbed expansion law $H(z)$, it is not clear that
they cannot affect the distance for redshifts where clustering has become relevant.
This is the point of view of workers on back reaction and clearly, before we have
not examined it in detail, we cannot exclude this possibility. Unfortunately, this is a 
relativistic effect of non linear clustering and our understanding of these effects 
is still rather poor.

Dyer \& Roeder, '72~\cite{DyRoe} have argued that the photons which end up in our
telescope go preferentially through empty or at least under-dense space and
therefore the distance formula should be corrected to the one of an open universe. 
But as we have seen, this is not sufficient and actually Weinberg, '76~\cite{Wein} has 
shown that the shear term which is present 
if matter is clustered in the case of simple 'Schwarzschild clumps' exactly corrects for the
missing Ricci term and reproduces the FL universe formula.
In a generic, clumpy spacetime the Sachs equation yields
$$ \frac{d^2 D_A}{dv^2} = -\left(\frac{1}{2}R_{\mu\nu}k^\mu k^\nu+|\si|^2  \right)D_A$$
Hence the presence of shear always leads to deceleration, like matter density. 
But the measured quantity is not $D_A(v)$ but $D_A(z)$ so we have to study
how the redshift is affected by clumping due to the motion of observers, $u^\mu$.
$$\frac{dz}{dv} = u_{a;b} k^a k^b $$
If the expansion of matter (observers) is substantially reduced in a clumping
universe this can reduce the redshift at fixed $v$ and therefore lead to seemingly
larger distance. Similar ideas have been put forward by Wiltshire~\cite{Wiltshire},
but of course we need to study this quantitatively. A quantitative study
the effects of structure formation on the distance-redshift relation has been attempted
by R\"as\"anen~\cite{syksy,syksy2}

Another possibility may be that the Universe is also statistically not homogeneous. 
From CMB observations we infer that it is very isotropic and this leaves us with 
spherically symmetric Lema\^\i tre-Tolman-Bondi (LTB) models. Clearly, this possibility
which violates 
the cosmological principle is not very attractive. It is therefore important to investigate 
whether we can test it observationally, and the answer is fortunately affirmative:  The relation 
between the speed of expansion, $H(z)$, and the distance $D_L(z)$ in an FL universe 
is given by Eq.~(\ref{e:DL}). In an LTB model this relation  no longer holds. Therefore
independent measurements of  both $H(z)$ and $D_L(z)$ (or $D_A(z)$) which test 
relation~(\ref{e:DL}), can check whether distances are really given by the FL expressions.
This at the same time also checks whether clustering modifies distances in an important way.
At present we do have relatively good distance measures out to $z\sim 0.5$ but no
independent measurements of $H(z)$. These may be obtained in the future from large 
galaxy surveys like DES or Euclid, which will allow us to measure separately the radial 
and the transverse matter power spectrum.

Other tests whether 'dark energy' is truly a new component in the stress energy tensor or
simply a misinterpretation of the observed distance can come from measurements of 
the growth factor of linear perturbations which we can determine with future weak lensing 
surveys like Euclid or via correlations of large scale structure and the integrated 
Sachs Wolfe effect. In a $\La$-dominated universe the linear growth of clustering is 
modified in a very specific way and  we would not expect a simple misinterpretation of 
observed distances to mimic also this behavior.
 
\section{Conclusions}\label{s:conc}
In this work I have pointed out that all {\em present} claims about the existence of dark energy have not measured $\Om_\La$ or even less $\Om_{DE}$ and $w$ directly, but just the distance 
redshift relation $D_L(z)$. They then have inferred the existence of dark energy by 
assuming the form~(\ref{e:DL}) for this relation, which holds in a FL universe. Even 
though many of you (especially the observers, I guess) may regard this point 
as trivial, I find it important to be aware of it before one is ready to postulate unobserved
scalar fields with most unusual properties, or violations of General Relativity on large scales.

I have not discussed the many possible pitfalls of the observations, which weaken 
any one observation, but my confidence relies on the fact that independent 
observations with different systematics find the same result.
I hope they are not too strongly  influenced by 'sociology', i.e.: if your finding disagrees 
with the results of others it 
must be wrong and therefore you do not publish it, however, if it agrees well it must be right and 
therefore you do not have to investigate every possible systematics which would increase 
your error bars and make your result less "competitive".

The beauty of research in cosmology is that data come in fast and there is justified 
hope that the question whether relation (\ref{e:DL}) holds for the real Universe, 
will be answered in the not very far future.
\vspace{2mm}

\noindent
{\bf Acknowledgement:}
I thank the organisors of the meeting for inviting me. I enjoyed many interesting 
and insightful discussions with many of the participants, especially with Roy Maartens.  
I am also grateful for discussions with Syksy R\"as\"anen and with Enea Di Dio
who pointed out an error in a first version.
This work is supported by the Swiss National Science Foundation and by the British 
Royal Society.

\end{document}